\documentclass[prl,twocolumn,floatfix,showpacs,superscriptaddress]{revtex4}
\usepackage{graphics}
\usepackage{epsfig}
\usepackage{times}
\usepackage{bm}

\begin{document}
\preprint{HEP/123-qed}
\title{Quantum electric dipoles in Dimer Mott insulator coupled to spin degrees of freedom}
\author{Chisa Hotta}
\affiliation{Department of Physics, Faculty of Science, Kyoto Sangyo University, Kyoto 603-8555, Japan}

\date{\today}
\begin{abstract}
We present an effective dipolar-spin model based on the strong coupling analysis, 
which may explain the possible origin of "spin liquid insulator". 
The issue is related to a dimer Mott insulator reminiscent of an 
organic triangular lattice system, $\kappa$-ET$_2$Cu$_2$(CN)$_3$, 
whose gapless spin liquid state has been discussed 
in the context of geometrical frustration of exchange coupling, $J$, between spins on dimer orbitals. 
It turns out that another degrees of freedom within the insulator, {\it quantum electric-dipoles} on dimers, 
interact with each other and significantly suppress $J$ 
through the dipolar-spin coupling, resulting in a possible {\it "dipolar-spin liquid"}. 
\end{abstract}
\pacs{75.40.-s, 72.80.Ng, 72.80.Le, 75.50.Mm, 77.22.Ej}   
\maketitle
\narrowtext 
A well known terminology "spin liquid" continues to attract attention 
ever since the Anderson's resonanting valence bond picture was presented\cite{anderson}. 
This state of matter is recently discussed in organic solid, 
$\kappa$-(BEDT-TTF)$_2$Cu$_2$(CN)$_3$\cite{shimizu03,syamashita08,myamashita09,ramirez08} 
(BEDT-TTF abbreviated as ET), 
in the context of absence of magnetic orders and of the possibility of gapless excitations. 
The solid has a triangular lattice structure at half-filling in unit of dimerized molecules, 
and is driven towards a Mott insulating state by the strong on-dimer Coulomb interaction\cite{kanoda}. 
The origin of its spin liquid state is then ascribed to the frustration effect from 
the nearly regular triangular geometry of the spin exchange coupling, $J'_{\rm eff}/J_{\rm eff} \sim 1$, 
which is estimated from the corresponding effective transfer integrals between dimer orbitals, 
$t'_{\rm eff}/t_{\rm eff} \!\sim \!1.05$\cite{komatsu96}. 
Theoretical works are developed successively within the half-filled single band Hubbard model (SBH) 
on the triangular lattice\cite{imada02,koretsune07,yoshioka09,kyung06,clay08}; 
an exotic non-magnetic (gapless) insulator is found in the anisotropy range of 
$t'_{\rm eff}/t_{\rm eff} \sim$ 0.7-1\cite{imada02}. 
Some interpretations are given by many-body spin exchanges beyond the Heisenberg model 
and the spinon Fermi surface\cite{motrunich05,senthil07} or by vison excitation\cite{sachdev09}. 
\par
Recently, however, $\kappa$-ET$_2$Cu$_2$(CN)$_3$ is reported to have 
anomaly at $T\!\!\sim$6K in lattice expansion coefficient\cite{lang09} as well as 
in dielectric constant which shows a relaxor ferroelectric-like behavior above this temperature \cite{majed09}. 
They suggest that the electronic degrees of freedom is still active in the insulating state. 
Coincidentally, model parameters are replaced from the above mentioned ones\cite{komatsu96} 
by the ab-initio calculation to $t'_{\rm eff}/t_{\rm eff}\!\sim \!0.8$\cite{nakamura09,valenti09}, 
$V/U_{\rm dimer}\!\sim$0.4 with $U_{\rm dimer}/t\!\sim \!15$\cite{nakamura09}, where $V$ and $U_{\rm dimer}$ 
are the inter-dimer and on-dimer Coulomb interactions, respectively. 
Therefore the geometrical frustration effect is not strong. 
Instead, the inter-site Coulomb interaction, $V_{ij}$, is large which shall play certain role in 
the low energy physics of this intriguing state. 
Since the system is quarter-filled in unit of molecule, there is an instability towards charge order 
by $V_{ij}$, which may compete with the Mott insulator\cite{chemrev}. 
In this Letter, we explicitly include $V_{ij}$ for the first time 
in describing $\kappa$-ET$_2$Cu$_2$(CN)$_3$ 
and demonstrate another scenario for the suppression of magnetic orders. 
We describe charge degrees of freedom in the insulating state 
as {\it "quantum electric-dipoles"} which fluctuate within the dimer. 
Dipoles couple to spins through the inter-dimer fluctuation of charges. 
The suppression of dipolar fluctuation by $V_{ij}$ leads to significant decrease of $J$'s and of magnetic correlations. 
\par
%
%
\par
We go back to the basic model of organic solids in unit of molecule \cite{chemrev}, 
a quarter-filled two-band extended Hubbard model, 
whose Hamiltonian is given as, 
\begin{equation}
{\cal H} = \!\!\!\!
\sum_{\langle i,j \rangle}\bigg(\sum_{\sigma=\uparrow\downarrow} 
\!\!t_{ij}\left(c^\dagger_{i\sigma} c^{\vphantom{\dagger}}_{j\sigma} + {\rm H.c.}\right) 
+ V_{ij} n_i n_j\bigg) + \sum_i U n_{i\uparrow}n_{i\downarrow}. 
\label{exthham}
\end{equation}
Here, $c^{\dagger}_{j\sigma}$ / $c^{\vphantom{\dagger}}_{j\sigma}$ 
are creation/annihilation operators of electrons with spin $\sigma$(=$\uparrow,\downarrow$) 
 and $n_{j\sigma}=c^\dagger_{j\sigma} c^{\vphantom{\dagger}}_{j\sigma}$ 
and $n_j=n_{i\uparrow}+n_{i\downarrow}$ are number operators. 
The model includes on-site ($U$) and nearest neighbor ($V_{ij}$) interactions. 
We consider strong dimerization effect, namely, each pair of sites connected 
by strong interactions, $(t_{ij},V_{ij})=(t_d,V_d)$, called "dimers" has one electron 
(i.e., half-filled in unit of dimer) on an average. 
\par
The half-filled SBH in Refs.\cite{imada02,koretsune07,yoshioka09,kyung06,clay08} 
is a limiting case of quarter-filled Eq.(\ref{exthham}), $t_d \gg t_{ij}$ and $V_{ij}=0$; 
In Eq.(\ref{exthham}) each dimer has sixteen bases which is reduced to four bases in SBH by the "dimer approximation". 
This four bases could account for charge fluctuation between dimers and describe both metal and dimer Mott insulator. 
Whereas, details of the two dimerized sites are neglected, 
e.g., a doubly occupied basis of SBH, $\big(\!\!\uparrow\downarrow\!\!\big)$, 
represents only one bonding state among six doubly occupied states of Eq.(\ref{exthham}), 
$(\uparrow\downarrow,0), (0,\uparrow\downarrow), (\uparrow,\downarrow), (\downarrow,\uparrow), 
(\uparrow,\uparrow),(\downarrow,\downarrow)$, 
where $\sigma$/0 denote the presence/absence of spin-$\sigma$ electrons on (site-1,site-2).  
The {\it intra-dimer charge disproportionation/fluctuation is not considered}, 
which is no longer legitimated when $V_{ij}$ is large ($\gtrsim t_d$)\cite{nakamura09}. 
%
\begin{figure*}[t]
\begin{center}
\includegraphics[width=16cm]{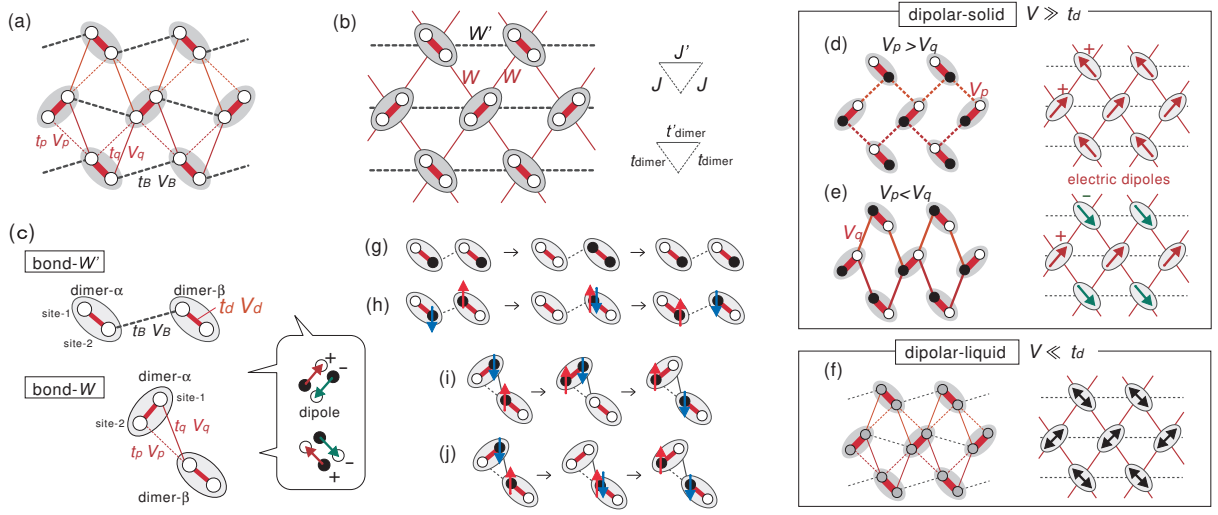}
\end{center}
\caption{(Color online) 
(a) $\kappa$-type lattice structure adopted to Eq.(1) in unit of molecule (circle); Bold line represents dimer bonds, $(t_d,V_d)$. 
(b) Anisotropic triangular lattice in unit of dimer, transformed from panel (a). 
(c) Two different types of connections between dimers in panel (a). 
Polarized (d)(e) and unpolarized (f) configurations of quantum dipoles. 
Charges avoid neighboring alignment of strong $V_{ij}$  
shown explictly in the fermionic representation (left panels). 
Panels (g)--(j) are the representative second order perturbation processes, which generate 
effective interactions in Eq.(\ref{ham2}), 
where filled circle and arrows represent the electrons and spins, respectively. 
}
\label{f1}
\end{figure*}
\par
Instead, we take the strong coupling limit (insulator) of Eq.(\ref{exthham}) as $t_{ij}/U, 
t_{ij}/V_d, V_{ij}/U, V_{ij}/V_d \rightarrow 0$. 
This approximation projects out bases having double occupancy on dimers with energy of order $U$ or $V_d$, and describes 
{\it both dimer Mott and charge ordered insulators}. 
The remaining bases keep exactly one fermion per dimer as, 
(site-1, site-2)=$(\uparrow,0), (0,\uparrow), (\downarrow,0)$, and $(0,\downarrow)$. 
We introduce electric-dipolar and spin operators, $\hat{P}$ and $\hat{S}$. 
Then, the above four states are described in the 2$\otimes$2 spinor representation as, 
$(P^z,S^z)=(\frac{1}{2},\frac{1}{2}),(-\frac{1}{2},\frac{1}{2}),(\frac{1}{2},-\frac{1}{2})$,
and $(-\frac{1}{2},-\frac{1}{2})$. 
The quantization $z$-axis of dipoles is fixed to the dimer-bond direction in real space. 
In the following, we derive the effective Hamiltonian 
by treating $t_{ij}$ and $V_{ij}$ ($\ne V_d$) perturbatively, 
by which {\it the charge fluctuation between dimers} is taken into account. 
\par
For the realization of Eq.(\ref{exthham}) in the bulk system, 
we choose a model lattice of $\kappa$-type organic solids \cite{kino96} shown in Fig.~\ref{f1}(a); 
the inter-dimer interactions are $(t_{ij},V_{ij})=(t_p,V_p), (t_q,V_q), (t_B,V_B)$ along three different bonds. 
If we take dimer as a unit, this lattice is mapped to an anisotropic triangular lattice in Fig.~\ref{f1}(b), 
whose horizontal ($W'$) and diagonal ($W$) inter-dimer bonds originate from $(t_B,V_B)$ and $(t_p,V_p),(t_q,V_q)$ of Fig.~\ref{f1}(a), respectively. The effective Hamiltonian, 
$\displaystyle {\mathcal H}_{\rm eff}= {\mathcal H}^{(1)} +{\mathcal H}^{(2)}+{\mathcal H}^{(3)}+{\mathcal H}^{(4)}$, 
is generated in unit of dimer in Fig.~\ref{f1}(b). 
The first order Hamiltonian is, 
\begin{equation}
{\mathcal H}^{(1)} = \!\!\! \sum_{l\in {\rm bond}W,W'} \!\! W_0^l P_{\alpha}^z P_{\beta}^z \;+\; \sum_\gamma t_d (P_\gamma^+ + P_\gamma^-), 
\label{ham1}
\end{equation}
where $W_0'\!=\!V_B$ and $W_0=V_q-V_p$. 
Eq.(\ref{ham1}) is nothing but a transverse Ising model of dipoles, 
where a competition of correlation and local quantum fluctuation of dipoles is imbedded, 
namely the Ising interaction term $W_0^l$($\sim V_{ij}$) versus the transverse field $t_d$. 
Representative dipolar states are shown in Figs.~\ref{f1}(d)-\ref{f1}(f); 
when $t_d \ll V_{ij}$, there are two different spacial orders depending on the geometry of $V_{ij}$. 
At $t_d\gg V_{ij}$, dipoles fluctuate and stay spacially uniform. 
The former "dipolar-solid" corresponds to charge order and the latter "dipolar-liquid" to dimer Mott insulator. 
At this order, the spin degrees of freedom are fully degenerate. 
\par
Degeneracy of spins are lifted at the second order level. 
The second order terms for two bonds ($W^l=W'$ and $W$ \cite{perturb}) yields, 
\begin{eqnarray}
\hspace*{-8mm}
{\mathcal H}^{(2)}_{l}\hspace*{-2mm}&=&\!\!
-W_{\rm pp}^{l} \; P_\alpha^z P_\beta^z + \hat{W}_{\rm ss}^{l} S_\alpha\cdot S_\beta 
-W_{\rm ppss}^{l}\big(P_\alpha^z P_\beta^z\big)\big(S_\alpha\cdot S_\beta\big) 
\nonumber\\
&&\hspace*{-8mm}
+ \hat{W}_{\rm p}^{l}(P_\alpha^++P_\alpha^-) 
-\hat{W}_{\rm pss}^{l} (P_\alpha^++P_\alpha^-)\big(S_\alpha\cdot S_\beta\big). 
\label{ham2}
\end{eqnarray}
Let us first focus on bond-$W'$. 
The first $(P_\alpha^z P_\beta^z)$-term originates from the process in Fig.~\ref{f1}(g). 
Noteworthy is the emergence of a dipolar-spin coupling term, $(P_\alpha^z P_\beta^z)(S_\alpha\cdot\!S_\beta)$, 
which together with $\hat{W'}_{\rm ss}$-term originate from the process in Fig.~\ref{f1}(h); 
exchange of spins occurs only when fermions occupy nearest neighbor sites, 
i.e., when dipoles are antiferroelectric $(P_\alpha^z,P_\beta^z)\!\!=\!\!(-\frac{1}{2},\frac{1}{2})$. 
This four-body term reminds of the Kugel-Khomskii Hamiltonian 
discussed in manganites\cite{kugelkhom,oles97} in the context of orbital-spin coupling. 
\par
The diagonal bond-$W$ consists of two interactions ($p,q$), 
each generating terms in the same manner as bond-$W'$. 
In addition, there are processes shown in Figs.~\ref{f1}(i) and \ref{f1}(j); 
Dimers exchange their fermions through two connections $p$ and $q$ 
by ending up flipping $P^z_\alpha$, namely, fermion in dimer-$\alpha$ changes its site location. 
This generates the last two terms in Eq.(\ref{ham2}) only for bond-$W$, i.e., $W'_{\rm p}=W'_{\rm pss}=0$. 
\par
Spins follow dipoles through these second-order dipolar-spin coupling terms in Eq.(\ref{ham2}). 
Effective interactions between spins on neighboring dimers ($\alpha$, $\beta$) 
are evaluated from the expectation values of 
coefficient of $S_\alpha\cdot S_\beta$ of Eq.(\ref{ham2}) as, 
\begin{eqnarray}
J'&=& \langle\hat{W}'_{\rm ss}\rangle - W'_{\rm ppss} \langle P_\alpha^z P_\beta^z \rangle, 
\nonumber\\
J&=& \langle\hat{W}_{\rm ss}\rangle - W_{\rm ppss} \langle  P_\alpha^z P_\beta^z \rangle -\langle \hat{W}_{\rm pss} (P_\alpha^+ + P_\alpha^-) \rangle. 
\label{jeff}
\end{eqnarray}
\par
In order to examine the actual competition between $t_d$ and $V_{ij}$, 
we perform exact diagonalization on ${\mathcal H}_{\rm eff}$ in finite dimer clusters\cite{sizedep}. 
We adopt $(t_p,t_q)=(0.7,-0.25)$ in unit of $t_B=1$
which describes $\kappa$-ET$_2$Cu$_2$(NCS)$_3$\cite{komatsu96}, and take $(U,V_d)\!=\!(15,10)$. 
These values are interpreted to coefficients $W^l$ and $\hat{W^l}$ of ${\mathcal H}_{\rm eff}$. 
For the choice of $V_{ij}$, we take $V_q\!>\!0$ and $V_p\!=\!V_B\!=\!0$, concentrating on the 
type of dipolar solid given in Fig.~\ref{f1}(e), which is sufficient to clarify the essential physics 
of the competition of $V_{ij}$ and $t_d$. 
\par
\begin{figure}[tbp]
\begin{center}
\includegraphics[width=9cm]{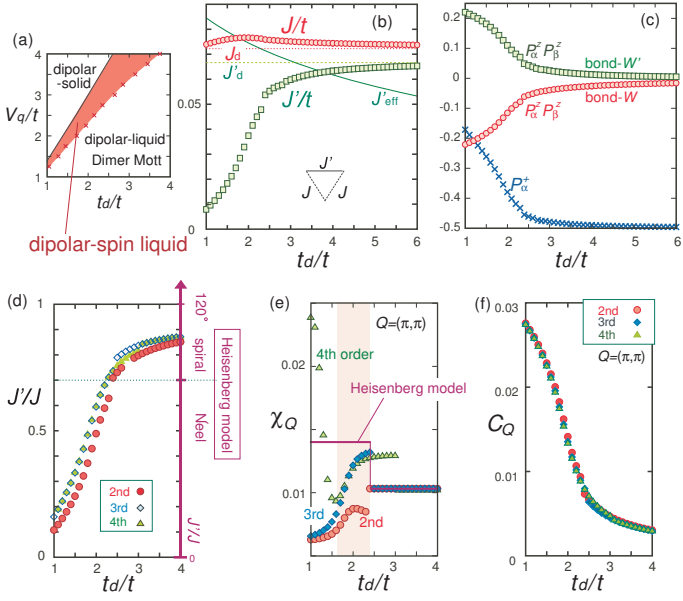}
\end{center}
\caption{(Color online) 
Results on ${\mathcal H}_{\rm eff}$ as functions of $t_d$ and $V_q$ with fixed $(t_B,t_p,t_q)\!=\!(1,0.7,-0.25)$ 
and $(U,V_d,V_p,V_B)\!=\!(15,10,0,0)$. 
(a) $t_d$-$V_q$ phase diagram. Along its $V_q=3$ line, 
(b) $J$ and $J'$, (c) $\langle P_\alpha^zP_\beta^z \rangle$ and $\langle P_\alpha^+\rangle$ are shown. 
The broken lines, $J_d$ and $J'_d$, are the $t_d\rightarrow \infty$ limit of $J$ and $J'$ 
obtained by taking $\langle P_\alpha^zP_\beta^z \rangle= 0$ and $\langle P_\alpha^\pm \rangle\!=\!-0.5$ in Eq.(\ref{jeff}). 
Solid line, $J_{\rm eff}'$, is the evaluation from the conventional dimer approximation. 
Panels (d)-(f) give the comparison up to second, third, and fourth order perturbation; 
(d) $J'/J$ and the structural factors of (e) spins $\chi_{Q}$ and (f) charges $C_Q$ with $Q=(\pi,\pi)$. 
The bulk ground state of Heisenberg model of the correpsonding $J'/J$ is shown on the right of panel (d). 
}
\label{f2}
\end{figure}

We first elucidate the phase diagram on the plane of $t_d$ and $V_q$ as shown in Fig.~\ref{f2}(a). 
As anticipated, the dipolar-solid (Fig.~\ref{f1}(e)) and liquid (Fig.~\ref{f1}(f)) appears at large $V_q$ and $t_d$, respectively. 
The solid-liquid phase boundary is determined as a minimum of charge gap of the $U\!=\!\infty$-limit of Eq.(\ref{exthham}). 
\par
Next, we vary $t_d$ along the fixed line of $V_q=3$ in the phase diagram at the second order level, 
${\mathcal H}_{\rm eff}= {\mathcal H}^{(1)} +{\mathcal H}^{(2)}$. 
Figure~\ref{f2}(b) shows that $J'$ is suppresssed in the dipolar-solid at small $t_d$, and increases significantly by $t_d$. 
This can be explained by the remarkable $t_d$-dependence of $\langle P_\alpha^zP_\beta^z \rangle$ 
shown in Fig.~\ref{f2}(c); 
In Eq.(\ref{jeff}), the second term of $J'$ with the constant coefficient, 
$W_{\rm ppss}'\!=\!\frac{4t\,^2}{U-V}\!>\!0$, 
has large negative contribution to $J'$ by $\langle P_\alpha^zP_\beta^z \rangle\!\simeq\! 0.25$ at $t_d \sim 1$, 
which goes to zero as $\langle P_\alpha^zP_\beta^z \rangle \rightarrow 0$ at $t_d \rightarrow \infty$. 
Similar discussion holds for $J$; 
the second and third term of Eq.(\ref{jeff}) are both positive but are decreasing and increasing functions of $t_d$, 
respectively, since $W_{\rm ppss}\!=\!\frac{4t_p\,^2}{U-V_p}\!-\!\frac{4t_q\,^2}{U-V_q}\!>\!0$, 
$\langle P_\alpha^zP_\beta^z \rangle\!\simeq\! -0.25 \!\rightarrow\!0$, 
and $W_{\rm pss}>0$, $\langle P_\alpha^\pm\rangle \!\simeq\! 0\!\rightarrow \!-0.5$. 
The two variations cancel out, keeping $J$ almost unchanged by $t_d$. 
\par
%
One important point is the unexpected decrease of $J'$ due to $V_{ij}$.  
In general, $V_{ij}$ works to screen $U$ and to enhance $J$\cite{brick95}. 
This effect actually appears in the increase of denominator of $W_{\rm ppss}\propto (U\!-\!V_{ij})^{-1}$ by $V_{ij}$. 
However, in our case $V_{ij}$ works directly on dipoles at the first order level, 
and varies $\langle P_\alpha^zP_\beta^z \rangle$ more significantly than their coefficients, suppressing $J'$. 
By contrast, $J'_{\rm eff}\!=\!4t'_{\rm eff}\,^2/U_{\rm dimer}$ obtained by the conventional dimer approximation based on SBH 
shows a "screening effect" as shown in Fig.~\ref{f2}(b), 
where $t'_{\rm eff}\!=\!\frac{t_B}{2}$, 
and $U_{\rm dimer}=2t_d+\frac{U+V_d}{2}- \frac{U-V_d}{2}\big(1+\big(\frac{4t_d}{U-V_d}\big)^2\big)^{1/2}$. 
\par
Another interesting aspect is the strong correlation between $J$ and $J'$. 
In the dipolar-solid state, the dipolar vectors are bound to particular spacial directions, 
which contribute to both $J$ and $J'$ through $\langle P_\alpha^zP_\beta^z \rangle\sim \pm 0.25$. 
The antiparallel dipoles along bond-$W'$ always suppresses $J'$, 
whereas along bond-$W$, the antiparallel dipoles in Fig.~\ref{f1}(d) suppress $J$ but parallel ones in Fig.\ref{f1}(e) do not. 
{\it In both cases}, however, we confirmed that  $J'/J$ is suppressed at small $t_d$ (see Fig.~\ref{f2}(d)). 
\par
In this way, at small $t_d$, the anisotropy of $J'/J$ becomes large and 
{\it the geometry of spin interactions approaches a square lattice}. 
For reference, we consider the bulk ground state of the Heisenberg model\cite{weihong99,trumper99,lhuillier92,sorella99}; 
at $J'/J \lesssim 0.7$ the antiferromagnetic order sets in, which is reflected in the 
spin structural factor, $\chi_{\bm Q}$, as a single peak at $\bm Q=(\pi,\pi)$. 
Figure~\ref{f2}(e) shows the peak amplitude of $\chi_{\bm Q}$ of our model versus the Heisenberg model calculated on the same finite cluster. 
The phase transition of the spin sector takes place at $J'/J\sim 0.7$, i.e., at $(t_d/t)_c \simeq 2.4$. 
At $t_d/t > (t_d/t)_c$ well inside the dipolar-liquid phase, the present model has the same $\chi_{\bm Q}$ as the Heisenberg one. 
By contrast, $\chi_{\bm Q}$ is significantly suppressed from that of the Heisenberg one at $t_d/t\lesssim (t_d/t)_c$. 
This indicates that in approaching the dipolar-solid phase, 
{\it the dimer spin degrees of freedom even at second order is no longer mapped onto the Heisenberg model}. 
In such case, the multiple spin-exchanges which emerge as higher order perturbation terms may become important. 
Therefore, we calculate all the perturbation processes up to fourth order 
$\mathcal H_{\rm eff}=\mathcal H^{(1)}+\mathcal H^{2)}+\mathcal H^{(3)}+\mathcal H^{(4)}$. 
(up to eight-body interactions, and including the four-dimer ring exchanges\cite{misguish98} which maximally amount to $J_4/t \sim 0.0027$). 
The numerical results up to third and fourth orders are compared with the second order ones in Fig.~\ref{f2}(d)-(f); 
$J'/J$ as well as the structural factor of dipoles ($C_{\bm Q}$ in Fig.~\ref{f2}(f)) are almost unchanged. 
By contrast, $\chi_Q$ shows a significant upturn ($t_d/t \lesssim 1.7$) at the fourth order level\cite{fourth}. 
Still, in the vicinity of the dipolar-solid/liquid phase boundary, $t_d\!=\!1.7\!-\!2.4$, 
there remains a strong suppression of antiferromagnetic correlation at all orders. 
The corresponding region is descibed as "dipolar-spin liquid" phase in Fig.~\ref{f2}(a). 
It apparently not originates from spin degrees of freedom alone. 
We consider that the liquid dipoles develop their short-range correlation, 
which rumple the distribution of charges, resulting in a spacially non-uniform value of $J$'s 
within the certain imaginary timescale, leading to the possible spin liquid state. 
Thus, in the dimer system, the four-body dipolar-spin interactions in Eq.(\ref{ham2}) play comparably important role as 
a well-known higher order ring exchange terms\cite{misguish98} in liquidizing the spin sector. 
\par
To summarize, we derived an effective dipolar-spin model relevant 
in the insulating phases of the quarter-filled extended Hubbard model with lattice dimerization. 
The intra-dimer charge degrees of freedom is described for the first time as quantum electric dipoles fluctuating by $t_d$ (the inter-dimer transfer integral). 
At large $t_d$ the conventional dimer Mott insulator, namely a dipolar-liquid, is stable. 
The inter-dimer Coulomb interaction, $V$, competes with $t_d$ and at large $V/t_d$ a dipolar-solid (charge order) emerge. 
The spins couple to the dipoles through the inter-dimer charge fluctuation, 
and in approaching the dipolar-solid from a liquid phase, strong modicication of spin-spin interactions, $J$ and $J'$, 
as well as significant suppression of the antiferromagnetic correlation is observed, 
indicating the existence of a {\it"dipolar-spin liquid"}. 
\par
We anticipate that the so-called spin liquid Mott insulator $\kappa$-ET$_2$Cu$_2$(NCS)$_3$ lies 
in our {\it"dipolar-spin liquid"} phase at low temperature, where both spins and charges remain short range ordered. 
The final comment is given on the comparison of parameters with experiments and other theories. 
$\kappa$-ET$_2$Cu$_2$(NCS)$_3$ undergoes a metal-insulator (MI) transition at $\sim$0.4GPa\cite{shimizu03}. 
The ab-initio calculation shows that $t_{ij}$ varies by 1.3 times from ambient pressure to 0.75GPa\cite{valenti09}, 
which brings $U_{\rm dimer}/t_{\rm eff}$ from 15 to 11\cite{nakamura09,pressure}. 
This value is already larger than the conventional empirical estimate of $U_{\rm dimer}/t_{\rm eff}\!\sim\!7\!-\!8$\cite{shimizu03}. 
In such case, our perturbative treatment up to fourth order can be qualitatively adopted, at least at ambient pressure. 
Starting from the same $t_{ij}$\cite{komatsu96} the conventional dimer approximation on a SBH 
reaches the effective Heisenberg coupling, $J_{\rm eff}'/J_{\rm eff}\sim 1$, 
whereas our {\it dipolar-spin-liquid} always has $J'/J\!\sim \!0.6\!-\!0.7$. 
Interestingly, our $J'/J$ coincides without assumption into a result of the ab-initio calculations as, 
$\sqrt{J'/J}\!\sim \!t_{\rm eff}'/t_{\rm eff}\! \sim \!0.8$\cite{nakamura09,valenti09}. 
Thus, we interpret the ab-initio results as a renormalized value of our effective $J$'s after including $V_{ij}$. 
Regarding the interpretation of 6K anomaly, 
we expect the following senario; at high tempearture ($T$), the thermal flucuation of dipoles is dominant (dipolar liquid is stable). 
At $T\sim 0$, the dipoles remain short range ordered as well due to the quantum fluctuation ($t_d$), 
and so as the spins which follow dipoles. 
In lowering $T$, the dipolar(electronic) correlation once grows, but still remains a liquid towards $T=0$. 
$T=6K$ is possibly a maximally correlated point (maximum $V/t_d$), where $J'/J$ is suppressed at most. 
In fact, a relaxor ferroelectric behavior of dielectric constant indicates a pseudo-transition at $T_c$=6K\cite{majed09}. 
The lattice anisotropy also takes a local maximum here, which implies that $J'/J$ also pass through an extreme value. 
To confirm this senario, further development in both theories and experiments are required. 
Thus, so far, one cannot exclude the possibility that the "gapless spin liquid" may not be the result of 
the geometrical frustration but of a strong correlation between spins and charges(dipoles). 
\par
We thank T. Sasaki, I. Terasaki, and M. Imada for discussions. 
This work is supported by Grant-in-Aid for Scientific 
Research (No.21110522, 19740218, 22014014) from the Ministry of Education, Science, Sports and Culture of Japan. 


\begin{thebibliography}{99}
\bibitem{perturb}Coefficients $W$ and $\hat{W}$ are of order $t_{ij}^2/M$ ($M=U,V_d,U-V_{ij}$), and $\hat{W}$ include $P^z_j$-term. 
The denominator includes $V_{ij}$ from surrounding dimers, 
while the modification is small at $U,V_d \gg V_{ij}$ and we neglect them for simplicity.
\bibitem{sizedep}Size dependence of $J$ and $J'$ is less than $2\%$ among 
$2\times 4, 3\times 4, 4\times 3$-clusters in Fig.\protect\ref{f1}(b). We adopt $3\times 4$ for Fig.\protect\ref{f2}.
\bibitem{fourth} The upturn of $\chi_Q$ originates from Heisenberg term, $J$, which appears at fourth order when $t_d$ is small. 
The second order is absent since charges localized(ordered) on every other site.
\end{thebibliography}
\end{document}